\documentclass[fleqn,usenatbib]{mnras}

\usepackage{newtxtext,newtxmath}
\usepackage{booktabs}	

\usepackage[T1]{fontenc}

\DeclareRobustCommand{\VAN}[3]{#2}
\let\VANthebibliography\thebibliography
\def\thebibliography{\DeclareRobustCommand{\VAN}[3]{##3}\VANthebibliography}

\usepackage{graphicx}	
\usepackage{amsmath}	
\usepackage{threeparttable}
\usepackage{booktabs} 
\usepackage[utf8]{inputenc}
\usepackage{subcaption}
\usepackage{fix-cm}
\usepackage[normalem]{ulem}
\usepackage{xcolor}
\usepackage{rotating}
\usepackage{adjustbox}
\usepackage{xcolor}

\newcommand\hi{$\textrm{H}\,\scriptstyle\mathrm{I}$}
\DeclareUnicodeCharacter{2212}{\textendash}

\title[Infalling gas in PKS 2014−55]{\hi{} gas in the rejuvenated radio galaxy PKS 2014-55}

\author[Leon K. Mtshweni et al.]{Leon K. Mtshweni$^{1}$\thanks{email:leon.kb.m.astro@gmail.com},
Kshitij Thorat$^{1}$, Roger P. Deane$^{1,2}$, Bradley S. Frank$^{4,5,6}$,  Filippo M. Maccagni$^{7}$, 
\newauthor Gyula I. Józsa$^{8,9}$, William D. Cotton$^{3,10}$, Gourab Giri$^{1,3}$, Sarah V. White$^{15}$, Marcellin Atemkeng$^{12}$, \newauthor Hertzog L. Bester$^{3,9}$, Bernie L. Fanaroff$^{3}$, Ian Heywood$^{13,9}$, Graham Lawrie$^{2}$, Thato E. Manamela$^{1}$, \newauthor Isaac Magolego$^{2}$, Tom Mauch$^{3}$, Nadeem Oozeer$^{3,9}$, Oleg Smirnov$^{3,9}$ and Masacheba S. Kupa$^{3}$
\\ \\
$^{1}$Department of Physics, University of Pretoria, Lynnwood Rd, Hatfield, Pretoria, 0002\\
$^{2}$School of Physics, University of the Witwatersrand, 1 Jan Smuts Ave, Braamfontein, Johannesburg, 2000\\
$^{3}$South African Radio Astronomy Observatory, 2 Fir Street, Black River Park, Observatory 7925, South Africa\\
$^{4}$STFC UK Astronomy Technology Centre, Royal Observatory, Edinburgh, Blackford Hill, Edinburgh EH9 3HJ, UK\\
$^{5}$Department of Astronomy, University of Cape Town, Private Bag X3, Rondebosch 7701, South Africa\\
$^{6}$The Inter-University Institute for Data Intensive Astronomy (IDIA), and University of Cape Town, Private Bag X3, Rondebosch 7701, South Africa\\
$^{7}$INAF – Osservatorio Astronomico di Cagliari, Via della Scienza 5, 09047 Selargius (CA), Italy\\
$^{8}$Max-Planck Institut fur Radioastronomie, Auf dem Hügel 69, 53121 Bonn, Germany\\
$^{9}$Department of Physics and Electronics, Rhodes University, P.O. Box 94 Makhanda 6140, South Africa\\
$^{10}$National Radio Astronomy Observatory, 520 Edgemont Road, Charlottesville, VA 22903, USA\\
$^{11}$International Centre for Radio Astronomy Research (ICRAR), Curtin University, Bentley, WA 6102, Australia\\
$^{12}$Department of Mathematics, Rhodes University, P.O. Box 94 Makhanda 6140, South Africa\\
$^{13}$Astrophysics, Department of Physics, University of Oxford, Keble Road, Oxford, OX1 3RH, UK\\
$^{15}$South African Astronomical Observatory, PO Box 9, Observatory, 7935, South Africa
}

\date{Accepted 2025 August 26. Received 2025 August 22; in original form 2024 September 19}

\pubyear{2024}

\begin{document}
\label{firstpage}
\pagerange{\pageref{firstpage}--\pageref{lastpage}}
\maketitle

\begin{abstract}
We present new high-spectral-resolution MeerKAT observations of \hi{} absorption against the central region of the restarted, giant, X-shaped radio galaxy PKS2014-55, which exhibits morphological evidence of three distinct cycles of activity. We report a wide component (FWHM 38 ± 7$\, $km$\text{s}^{-1}$) redshifted to 96 ± 50$\, $km$\text{s}^{-1}$, a deep-narrow detection (FWHM 19 ± 6$\, $km$\text{s}^{-1}$) which is redshifted to 160 ± 40$\, $km$\text{s}^{-1}$ and a shallow component (FWHM 22 ± 6$\, $km$\text{s}^{-1}$) redshifted to 240 ± 40$\, $km$\text{s}^{-1}$. One of the three components exceeds the typical rotational velocity of 100$\, $km$\text{s}^{-1}$, suggesting complex kinematics of the inflowing gas. These \hi{} observations support the correlation between the occurrence of \hi{} absorption and the rejuvenation of radio activity.
\end{abstract}

\begin{keywords}
galaxies: individual: PKS 2014-55 – galaxies: active – galaxies: ISM.
\end{keywords}
\section{Introduction}\label{intro_sect}
Large-scale radio jets from radio-loud Active Galactic Nuclei (RLAGN) offer a unique opportunity to study the AGN duty cycle through observations of remnant lobes from previous activity cycles. The triggering and fueling of such activity are closely linked to the mode of accretion onto the central black hole, which varies across AGN types. High-excitation radio galaxies (HERGs) are associated with radiatively efficient accretion, often driven by mergers or interactions, while low-excitation radio galaxies (LERGs) are typically fueled by hot halo gas or secular processes \citep{pierce_2022}. Deep optical imaging supports this distinction, revealing that most HERGs show morphological signs of past interactions, with over 94\% of strong-line radio galaxies exhibiting features such as shells, tails, or dust lanes \citep{2011Almeida}. Broad-band radio observations help constrain the spectral energy distributions of the lobes, offering estimates of timescales for both the active and quiescent phases of AGN activity \citep{Brienza2021}. A well-known subclass of episodic sources are the double-double radio galaxies (DDRGs), which display two distinct pairs of lobes, typically aligned along the same axis \citep{Schoenmakers2000}. There are also now known cases of the so-called Triple-Double radio galaxies (TDRGs) \citep{Chavan_tdgr}, in which three distinct epochs of activity are seen. Estimates based on spectral ageing, lobe expansion models \citep{1987Alexander, 2008Shabala, 2013aKonar}, and numerical simulations \citep{1997Tucker, 2004Omma} suggest that the active phase can last tens of Myr, interrupted by quiescent periods of $\sim$1 Myr \citep{2013aKonar}. The recurrence of jet activity is widely attributed to renewed accretion from the interstellar or circumgalactic medium, which replenishes the central engine with fuel \citep{rees_1984}.\\\\
A growing number of studies \citep{2009AMorganti,2014Maccagni,2023Yu} have investigated redshifted atomic neutral Hydrogen (\hi{}) within the central regions of radio galaxies, observed as the 21\,cm absorption line. Redshifted \hi{} absorption has been interpreted as evidence of cold gas infall in several young and restarted radio galaxies, including 3C\,236, B2\,0258+35, and 4C\,29.30 \citep{2012Struve,2017Maccagni}. These systems illustrate how episodic accretion events can reignite AGN activity, with \hi{} serving as a powerful diagnostic of fuelling mechanisms. According to \citet{2015Gereb_zoo}, \hi{} line profiles can be classified into three groups based on their kinematic widths: (1) narrow profiles (FWHM $<$ 100\,km\,s$^{-1}$), often caused by rotating \hi{} disks or clouds; (2) intermediate profiles (100\,km\,s$^{-1}$ $<$ FWHM $<$ 200\,km\,s$^{-1}$), typically reflecting more complex kinematics; and (3) broad profiles (FWHM $>$ 200\,km\,s$^{-1}$), likely originating from disturbed gas in mergers or outflows. Redshifted components in recurrent AGN are particularly relevant, as they are frequently linked to the cold gas infall that sustains renewed AGN activity \citep{Saikia2007,Chandola2010,2010Salter}.\\ \\
This paper examines the recurrent activity of the 1.57 Mpc Giant, X-shaped Radio Galaxy (XRG) PKS 2014-55\footnote{This source is also known as G4Jy 1613 in the G4Jy Sample \citep{2020AWhite, 2020BWhite}}, recently observed using MeerKAT \citep{cotton_thorat}. The latter publication revealed an outer cocoon of diffuse, steep-spectrum relativistic plasma, interpreted as evidence of an earlier epoch of activity. Two previous epochs were identified in the Australia Telescope Compact Array (ATCA) observation \citep{2008Saripalli}, suggesting three distinct epochs of activity for PKS 2014-55, making it a rare example of a TDRG, XRG, and Giant Radio Galaxy (GRG). 
In this study, we present high spectral resolution MeerKAT observations of the neutral Hydrogen towards the core of PKS 2014-55.  The structure of this paper is as follows: First, we provide a brief description of some basic properties of PKS 2014-55 and its host PGC 064440 in Sect.~\ref{source_info}. Sect.~\ref{observations} covers the MeerKAT observations and the data reduction process. Our findings and their interpretation are presented in Sect.~\ref{absorption}, and Sect.~\ref{summary} provides a summary. Throughout this work, we assume a $\Lambda$CDM cosmology of a Universe with $H_{\circ} = 70\, $km$\, $s$^{-1}\, $Mpc$^{-1}\, $, $\Omega_{\Lambda} = 0.286$ and $\Omega_{m} = 0.714$, \cite{spergel_cosmo_consts} and the web-based cosmology calculator \citep{cosmo_calc}
\section{PKS 2014-55}\label{source_info}
PKS~2014$-$55 is a radio source associated with the galaxy PGC~064440 \citep{1989Paturel}, which has a spectroscopic redshift of $z = 0.060629 \pm 0.000150$ \citep{2004Jones}. This corresponds to a luminosity distance of 255~Mpc, for which 1~arcsec translates to 1.2~kpc. The total projected linear size of the radio structure is 1.5~Mpc, inferred from an end-to-end angular separation of $\sim$22~arcmin. In contrast, the low-surface-brightness backflow wings span $\sim$14~arcmin, corresponding to a linear extent of 0.98~Mpc (see Fig.~\ref{fig:total_intsense_sofia_continuum_contours}). These dimensions classify PKS~2014$-$55 as a Giant Radio Galaxy (GRG). At higher resolution, the compact radio core resolves into an FR-II type inner double, with a total angular separation of $\sim$27~arcsec, or 32~kpc in projected scale. A more detailed account of the group environment surrounding PGC~064440 is provided in \citet{cotton_thorat}.
\section{OBSERVATIONS}\label{observations}
MeerKAT observed PKS 2014-55 for a total on-target observation of 10 hours from 29 February 2020 01:47 to 29 February 2020 14:16 (UTC), using 61 antennas, for an on-source time of 10 hours. The bandwidth of 856 MHz was observed in the 32k mode (32768 channels), yielding a channel width of 0.26123 MHz for the L-band (900–1670 MHz) observation. The data points were smoothed to an 8$\, $s integration. The observing sequence alternated between J1939-6342 for 2 minutes and PKS 2015-55 for 13 minutes. All reductions were performed on the ilifu computing facility\footnote{\url{http://www.ilifu.ac.za.}}. Data were reduced using the $\bf{{\tt CARACal}}$ pipeline\footnote{\url{https://caracal.readthedocs.io}} \citep{caracal}. $\bf{{\tt CARACal}}$ is a containerized scripting framework based on $\bf{{\tt STIMELA}}$\footnote{\url{https://github.com/ratt-ru/Stimela-classic}}, a platform-independent radio interferometric data reduction software suite that allows users to run various open-source radio interferometric data reduction and imaging tools within a single script. We split out a 50 MHz chunk around the observed 21 cm line at the spectroscopic redshift 0.06 (redshifted to 1.34 GHz), producing a total of 1912 channels. \\
\begin{table}
   \caption[Observational Information]{MeerKAT observations.} 
   \label{tab:MeerKAT observations}
   \small 
   \centering 
   \begin{tabular}{ll} 
   \toprule[\heavyrulewidth]\toprule[\heavyrulewidth]
   Observation Date         & 29 February 2020\\
   Bandwidth                & 856$\, $MHz\\
   Channel width            & 26.123$\, $kHz\\
   Integration time         & 8$\, $seconds\\
   Number of antennas       & 61\\
   Bandpass/flux calibrator & J1939-6342\\
   Gain calibrator & J1939-6342\\
   Time on target           & 10$\, $h \\
   Time on Calibrator       & 2$\, $h \\
   Pointing centre (J2000)  & RA  20:18:01.26\\
                            & Dec -55.39.30.50\\
   \bottomrule[\heavyrulewidth] 
   \end{tabular}
\end{table}
\noindent
\subsection{Calibrator flagging and cross calibration}
Basic flagging was performed, using $\bf{{\tt CASA}}$ \citep{2022PASP..134k4501C} and $\bf{{\tt TRICOLOUR}}$ \citep{2022ASPC..532..541H} to flag the RFI based on the Stokes Q visibilities. We used $\bf{{\tt CRYSTALBALL}}$ \citep{2022ASPC..532..409S} to populate the $\bf{{\tt MODEL\_DATA}}$ column with a detailed, image-based calibrator sky model that includes the sky substructure and sources within the MeerKAT primary beam. This step is necessary due to the high sensitivity of the MeerKAT telescope. The time-dependent delay (K) and complex gain (G) solutions of each antenna were derived using the standard $\bf{{\tt CASA}}$ task $\bf{{\tt gaincal}}$, avoiding RFI by excluding baselines shorter than $150$ meters. Additionally, the $\bf{{\tt bandpass}}$ task within $\bf{{\tt CASA}}$ was utilised to determine the bandpass (B) solutions, accounting for frequency-dependent errors. The determination of these gains was done in two iterations, with a round of flagging in-between to improve the fidelity of the gain solutions. Finally, the solutions were applied with $\bf{{\tt applycal}}$ to both calibrator and target visibility data, for the latter using the OTF splitting capability provided by the CASA task $\bf{{\tt mstransform}}$, which we used to get a 50 MHz chunk around the systemic \hi{} line.
\subsection{Target calibration and Continuum Imaging}
The cross-calibrated target visibilities were then flagged using $\bf{{\tt TRICOLOUR}}$ to perform a shallow round of automated sum-threshold flagging, which is good enough for the 50 MHz chunk of data we are using due to a relatively clean RFI environment. We subsequently imaged the target data with $\bf{{\tt WSClean}}$ \citep{2014MNRAS.444..606O}, employing the multiscale cleaning algorithm \citep{2017MNRAS.471..301O} to effectively clean the diffuse emission in the field and a Briggs \textit{robust} -0.5, to achieve an optimal trade-off between SNR and resolution given our limited bandwidth and which also avoids substructure within the synthesised beam. $\bf{{\tt CUBICAL}}$ \citep{2018MNRAS.478.2399K} was used, with a solution interval of 60 seconds to self-calibrate delays, amplitudes and phases. The calibration loop consisted of two delay self-calibration rounds and one amplitude and phase self-calibration round. A clean mask was generated using $\bf{{\tt breizorro}}$\footnote{\url{https://github.com/ratt-ru/breizorro}}, which applies a minimum filter over 50×50 pixel windows to estimate the local background to identify significant emission. We used the default threshold of 6.5$\sigma$ to create a binary mask that reliably identifies emission. The resulting binary mask was applied during the final Multi-Frequency Synthesis (MFS) imaging of the self-calibrated visibilities. The final continuum image has an rms noise of 7.7 $\mu$Jy beam$^{-1}$, with a synthesised beam of 6.7$\arcsec \times$ 5.9$\arcsec$, and a position angle of 79.2 deg.
\subsection{Spectral line imaging} 
Continuum subtraction from the above image was initiated by first predicting the source list that was generated by $\bf{{\tt WSClean}}$ during imaging into the $\bf{{\tt MODEL\_DATA}}$ column, using $\bf{{\tt CRYSTALBALL}}$. This produces a high fidelity model with good spectral resolution and avoids the step-function seen in the model data created using $\bf{{\tt WSClean}}$.
The continuum was then removed from the data by subtracting the $\bf{{\tt MODEL\_DATA}}$ column from the $\bf{{\tt CORRECTED\_DATA}}$ column. To further eliminate any residual continuum, the $\bf{{\tt uvcontsub}}$ function within the $\bf{{\tt CASA}}$ task $\bf{{\tt mstransform}}$ was employed. $\bf{{\tt uvcontsub}}$ constructs a spectral model for the continuum by applying polynomial fits to both the real and imaginary parts throughout the spectral window. \\ \\
Using $\bf{{\tt WSClean}}$, we converged on an optimal data cube through a sequence of imaging runs, with a Briggs robustness parameter of +1.0, to achieve a balance between sensitivity and resolution. This process yielded a cube with an angular resolution of 24.8$\arcsec \times$ 19.7$\arcsec$ and a position angle of -64.9 deg. We used $\bf{{\tt SoFiA2}}$ (Source Finding Application) \citep{sofia2_serra, sofia2_westmeier} to reliably characterize the \hi{} emission in the field. 
A 3$\sigma$ rms noise threshold was applied in each smoothing step and a reliability threshold of 0.70 to reliably identify detections. The \hi{} masks produced through this method were utilised to create moment maps from the original \hi{} cube. The resulting outputs from $\bf{{\tt SoFiA2}}$ include: a moment zero map illustrating all the \hi{} emission detections (Fig.~\ref{fig:total_intsense_sofia_continuum_contours}). An analysis of these findings is presented in the next section.
\section{Results and Discussion}\label{absorption}
\subsection{\texorpdfstring{\hi{}}{HI} Absorption in PKS 2014-55}\label{absorption_sub}
Figure~\ref{fig:spectrum} presents the \hi{} absorption spectrum detected against the compact radio core of PKS~2014$-$55. The spectrum was extracted from the full-resolution cube, which has a channel width of 5.855 \,km\,s$^{-1}$ (26.123 kHz), and subsequently smoothed using a boxcar filter of width 3 channels, resulting in a resolution of 17.565  \,km\,s$^{-1}$ (78.369 kHz). The rms noise level in the spectrum is 23\,$\mu$Jy\,beam$^{-1}$\,channel$^{-1}$. At this sensitivity, we reach a 3$\sigma$ column density detection limit of $6.3 \times 10^{18}$\,cm$^{-2}$. Visual inspection reveals three distinct components in the absorption profile, which we modelled using Gaussian fits. The spectral features can be described as follows: a broad component with a FWHM of 38 ± 7\,km\,s$^{-1}$ redshifted by 96 ± 50\,km\,s$^{-1}$; a narrow-deep component with a FWHM of 19 ± 6\,km\,s$^{-1}$ redshifted by 160 ± 40\,km\,s$^{-1}$; and a shallow component with a FWHM of 22 ± 6\,km\,s$^{-1}$ redshifted by 240 ± 40\,km\,s$^{-1}$. All three components are redshifted relative to the systemic velocity of the host galaxy. Table~\ref{tab:summary_HI} summarises the derived physical parameters of the absorbing gas.\\ \\
This triple-component absorption profile places PKS~2014$-$55 among a small number of rejuvenated radio galaxies exhibiting redshifted \hi{} absorption \citep{Saikia2007,Chandola2010,2010Salter}. While redshifted features are often interpreted as signatures of cold gas infall toward the nucleus, the detection of three kinematically distinct components suggests a complex, structured reservoir of gas. The combination of a deep, narrow core and both broad and shallow features may indicate multiple streams of inflowing material, or a mix of dynamically settled and unsettled gas. \\ \\
We determine the optical depth using the following relation:
\begin{equation}
e^{-\tau_{\nu}} = 1 - \frac{S_{\nu}^{\text{\hi{}}}}{S_{\text{1.4\,GHz}}},
\label{eq:optical_depth}
\end{equation}
where $S_{\text{1.4\,GHz}}$ is the continuum flux density and $S_{\nu}^{\text{\hi{}}}$ is the observed line flux density at frequency $\nu$. The corresponding H\,\textsc{i} column density is calculated using:
\begin{equation}
N_{\text{\hi{}}} = 1.83 \times 10^{18} \frac{T_{\text{s}}}{f_{\text{c}}} \int \tau(\nu) \, d\nu \quad \text{[cm}^{-2}\text{]},
\label{eq:column_density}
\end{equation}
where $T_{\text{s}}$ is the spin temperature and $f_{\text{c}}$ is the covering factor. Assuming $T_{\text{s}} = 100$\,K and $f_{\text{c}} = 1$, the integral can be approximated as $\int \tau(\nu) \, d\nu \approx \tau \times \text{FWHM}$ for each component identified in Fig.~\ref{fig:spectrum}. This yields column densities of 2.6 ± 0.3 $\times 10^{20}$\,cm$^{-2}$, 3.0 ± 0.2 $\times 10^{20}$\,cm$^{-2}$, and 1.3 ± 0.1 $\times 10^{20}$\,cm$^{-2}$ for the broad, narrow-deep, and shallow components, respectively, using a background integrated continuum flux density of 69.0 ± 0.01\,mJy.\\ \\
The detection of multiple redshifted \hi{} absorption components against the radio core provides direct evidence of structured cold gas inflow in a system that lacks clear signatures of major mergers. Unlike other rejuvenated radio galaxies with redshifted \hi{} absorption, such as 3C~236 and 4C~29.30 \citep{2012Struve,2017Maccagni}. PKS~2014$-$55 also shows no evidence of fast outflows. This reinforces a scenario in which episodic AGN reactivation is driven by infall rather than feedback.\\ \\
We find no evidence of \hi{} absorption at any other location, nor do we detect fast AGN-driven outflows of cold gas within the 50\,MHz bandwidth, which spans a velocity range of 11\,000\,km\,s$^{-1}$ (1912 channels). \\ \\
Given our continuum flux density of 69.0 ± 0.01\,mJy and a spectral RMS of 23\,$\mu$Jy/beam per channel, we place a conservative 3$\sigma$ upper limit on the peak optical depth of $\tau \lesssim 0.001$ for any broad component with FWHM $\gtrsim 300$\,km\,s$^{-1}$. This effectively rules out the presence of strong, high-velocity outflows along our line of sight, unless the gas is either sufficiently diffuse or offset from the radio core to escape detection in absorption. \\ \\
Moreover, no \hi{} emission is detected from the source itself. This could either reflect a genuine scarcity of extended neutral gas or a limitation in sensitivity due to the bright underlying radio continuum. The presence of a tilted dust lane in the host galaxy PGC 064440 supports the idea that cold gas has been externally acquired, likely through a minor merger or accretion event \citep{dust_lane4}. Tilted dust lanes are in some instances interpreted as signs of externally supplied gas that has yet to settle into the stellar disk. Simulations show that such gas can gradually align with the host galaxy's plane over time \citep{2015Lagos}. While we cannot resolve the absorbing gas spatially, the presence of a dust lane implies that cold gas is present in the host galaxy. If this gas is settling inward, it is reasonable to consider that the absorption may originate near the AGN core. Given the angular resolution of MeerKAT, however, higher resolution observations would be needed to effectively localise the gas seen in absorption within the host galaxy.\\ \\
Blueshifted \hi{} outflows are frequently reported in young or restarted radio galaxies, particularly where compact jets interact with a dense, clumpy ISM \citep{2005AMorganti, 2018Morganti2, 2019Morganti}. The absence of any broad, blueshifted \hi{} absorption in PKS 2014-55 is noteworthy. Despite its restarted activity and compact radio structure, both conditions that often favour jet–ISM interaction, we observe only redshifted, multi-component absorption. These findings support a scenario in which cold gas inflow currently fuels the AGN, with no observational evidence for significant merger activity. The optical isophotes of the host galaxy, as seen in Figure 9 of \citet{cotton_thorat}, show little deviation from ellipticity and reveal no obvious substructure.\\ \\
While the occurrence of \hi{} in restarted AGN has been noted in a few systems \citep{Saikia2007, Chandola2010, 2010Salter}, the small sample size limits general conclusions. The detection of a redshifted, multi-component \hi{} absorption spectrum in PKS 2014-55 (Fig.~\ref{fig:spectrum}) offers valuable insight into the cold gas conditions in such rejuvenated systems.\\ \\
Although all three absorption components are redshifted, some of the gas could still originate from a rotating structure within the host galaxy. This remains plausible given that all components exhibit FWHM values well below 100$\, $km$\, $s$^{-1}$, a regime typically associated with large-scale rotating disks or clouds \citep{2015Gereb_zoo}. However, the shallowest component is characterised by a velocity offset of $240 \pm 40\, \text{km}\, \text{s}^{-1}$ from the systemic redshift. Such a large displacement suggests this gas is unlikely to be part of a stable rotating disk, and may instead represent a disturbed or infalling cloud.
\subsection{Environmental context}\label{emission_sub}
\begin{figure}
\centering
\includegraphics[width=0.5\textwidth]{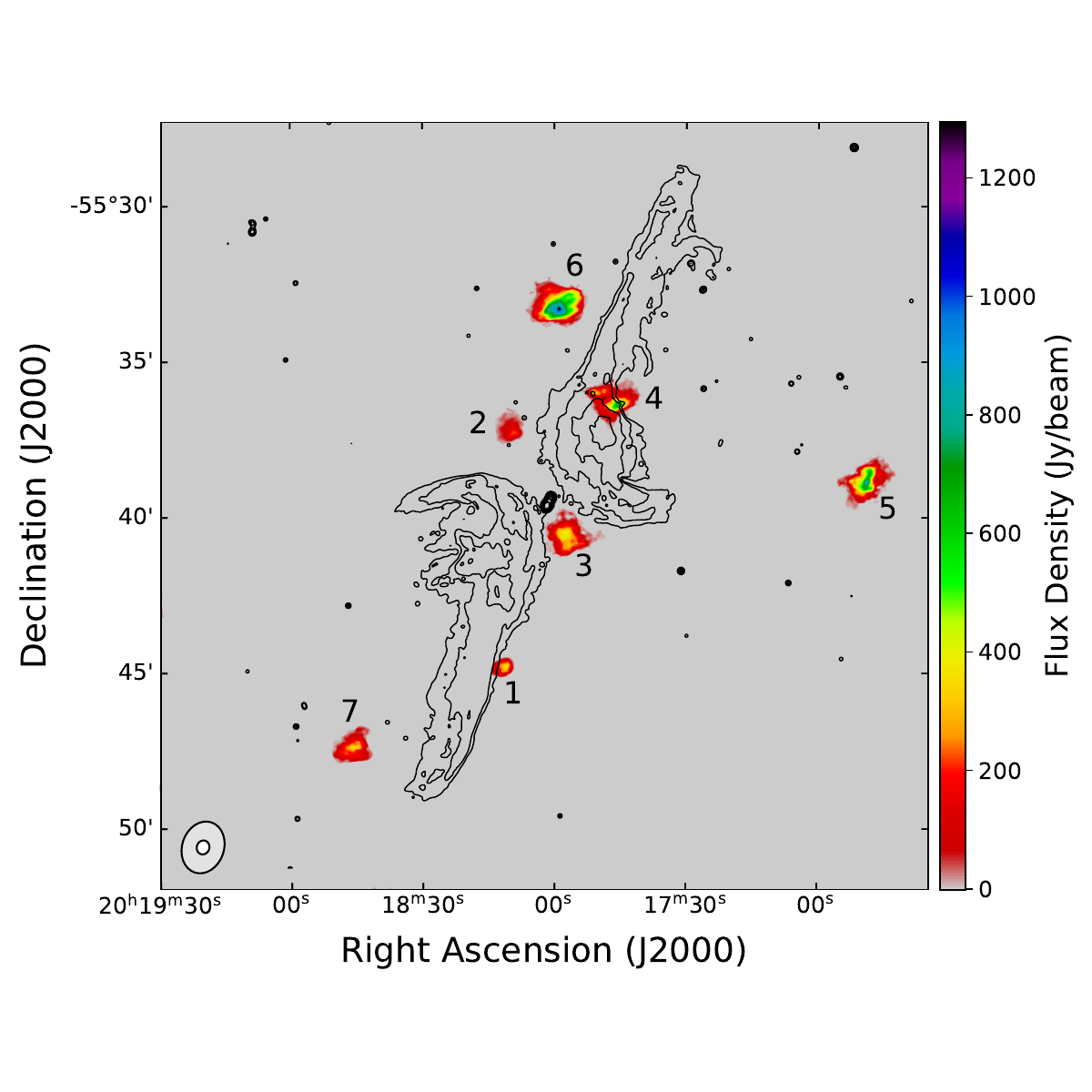}
\caption[Velocity\_field]{A total intensity map of the seven \hi{} emission sources covering 10 arcminutes around the optical host, PGC 064440. The radio continuum contours of PKS 2014-55 are overlaid at levels of 100$\, \rm{\mu Jy/beam}\, \times 2^1, 2^2,...,2^{10}$ \citep{cotton_thorat}. The continuum map's restoring beam has dimensions of 7.4$\arcsec \times$ 7.4$\arcsec$, with a position angle of 89.9 deg, represented by the black outlined circle in the bottom left. The moment zero map's restoring beam is sized 24.8$\arcsec \times$ 19.7$\arcsec$, with a position angle of -64.9 deg, denoted by the white outlined circle in the bottom left.} 
\label{fig:total_intsense_sofia_continuum_contours}
\end{figure}

\begin{figure}
\centering
\includegraphics[width=0.5\textwidth]{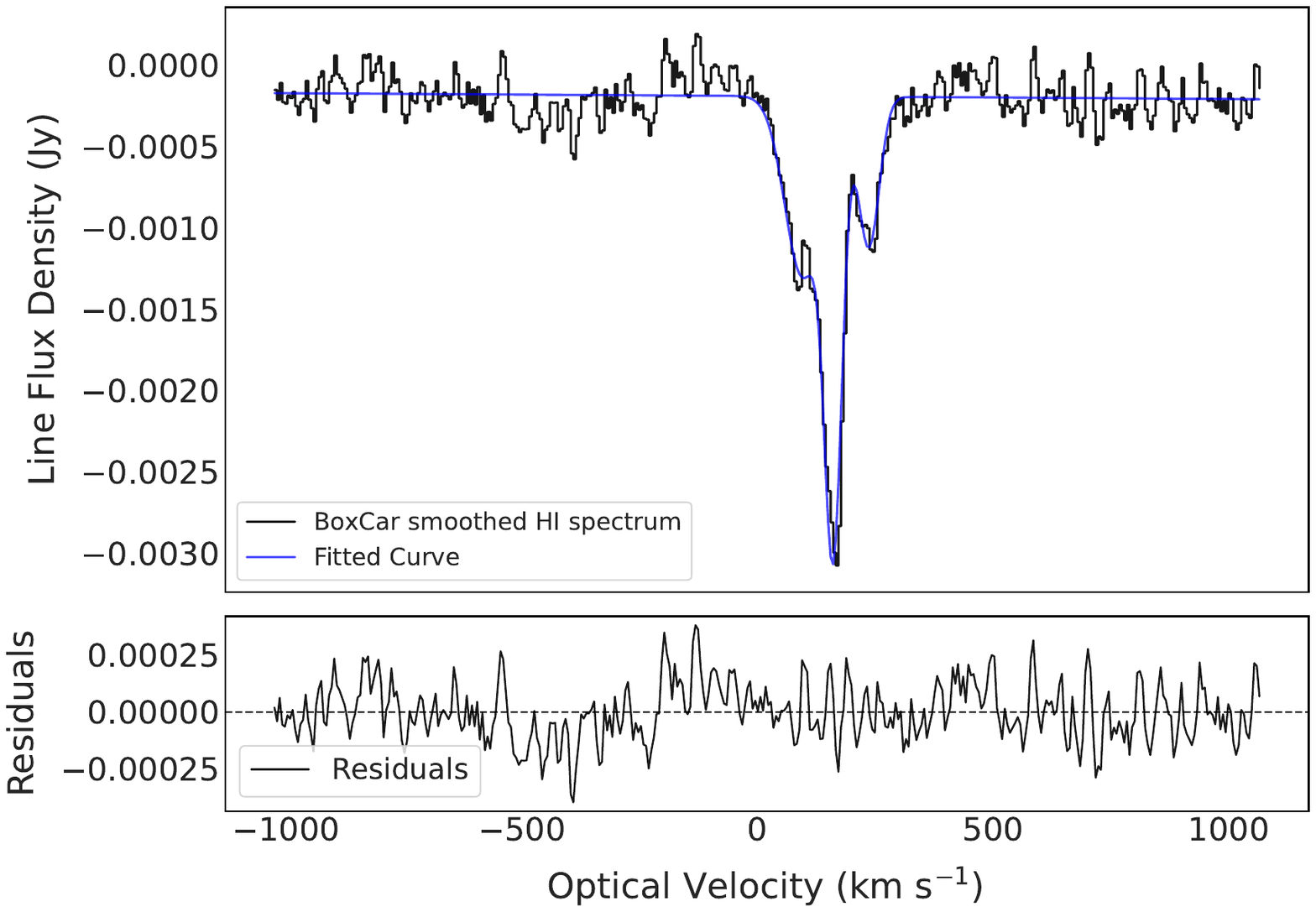}
\caption[Double Gaussian Fit to \hi{} Absorption Spectrum]{High-resolution \hi{} absorption spectrum toward the core of PKS 2014-55, extracted at a channel width of 5.855 \,km\,s$^{-1}$ and enhanced using a 3-channel boxcar filter. The data have a restoring beam of $24.8\arcsec \times 19.7\arcsec$ at a position angle of $-64.9^{\circ}$. The spectrum is velocity-offset to the systemic optical redshift and is well described by a three-component Gaussian model. Residuals from the fit are shown in the bottom panel.}
\label{fig:spectrum}
\end{figure}

\begin{figure*}
\centering
\begin{tabular}{cc}
  \begin{subfigure}{0.45\textwidth}
    \includegraphics[width=\linewidth]{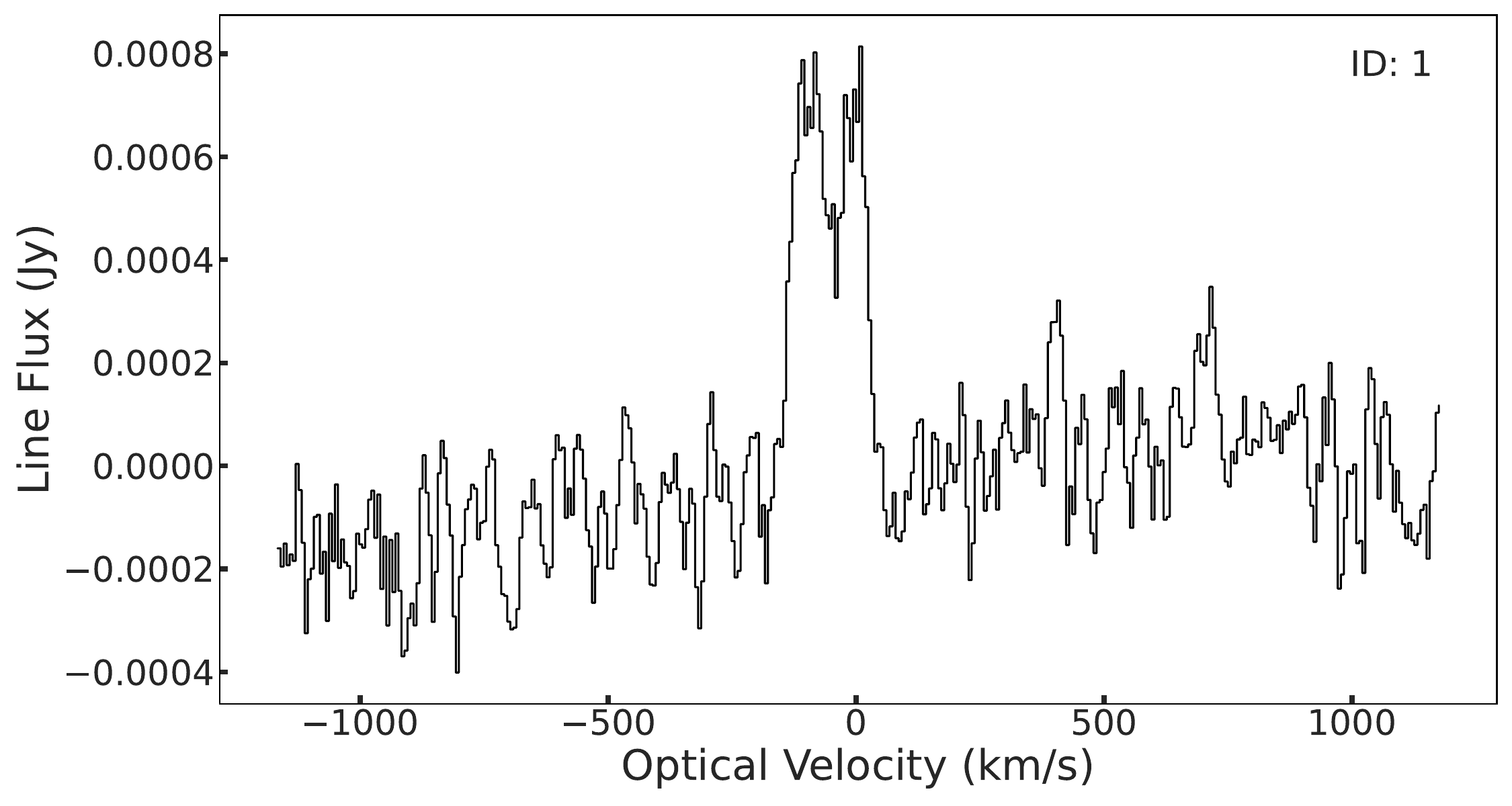}
  \end{subfigure} &
  
  \begin{subfigure}{0.45\textwidth}
    \includegraphics[width=\linewidth]{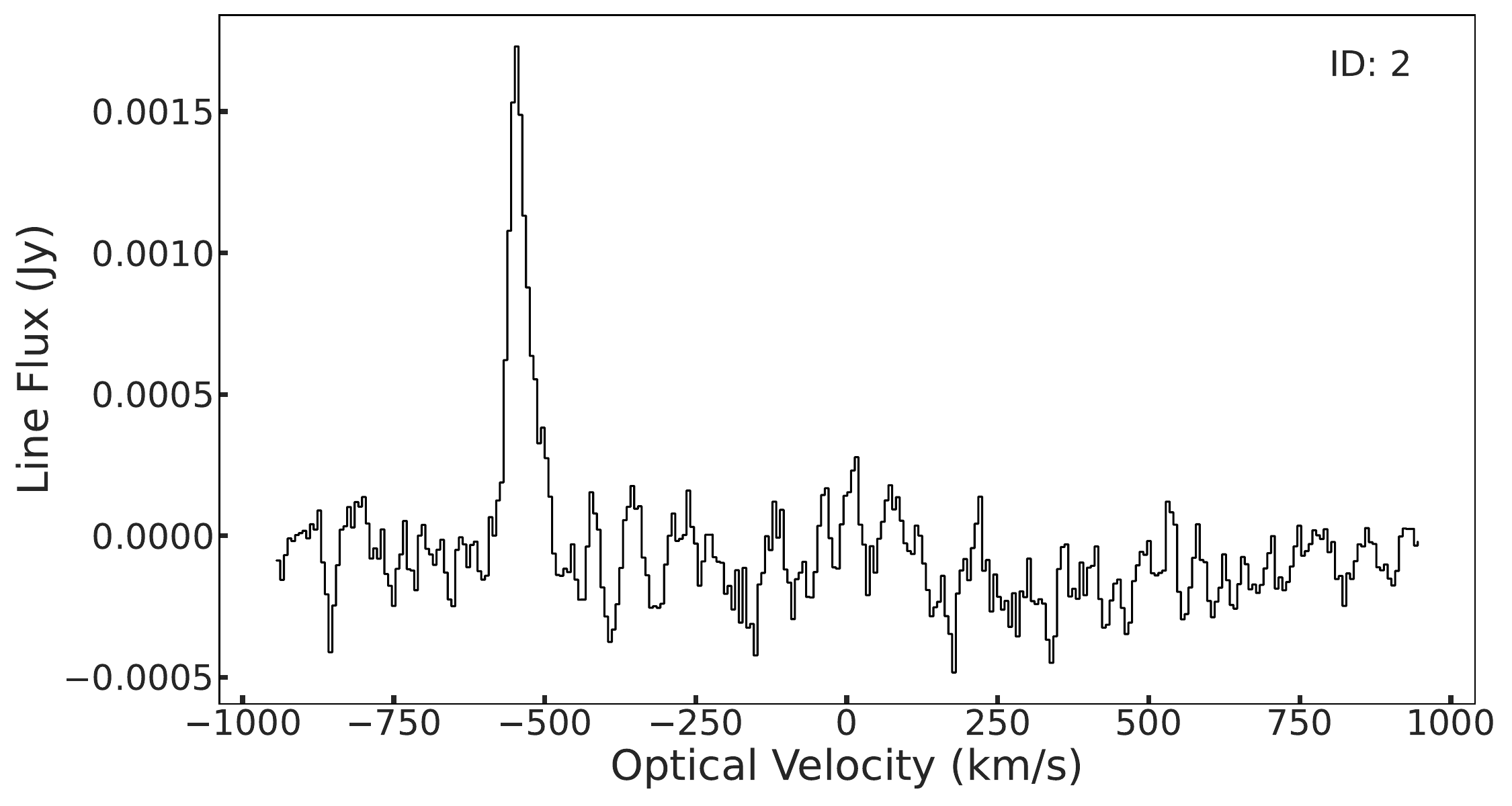}
  \end{subfigure} \\
  
  \begin{subfigure}{0.45\textwidth}
    \includegraphics[width=\linewidth]{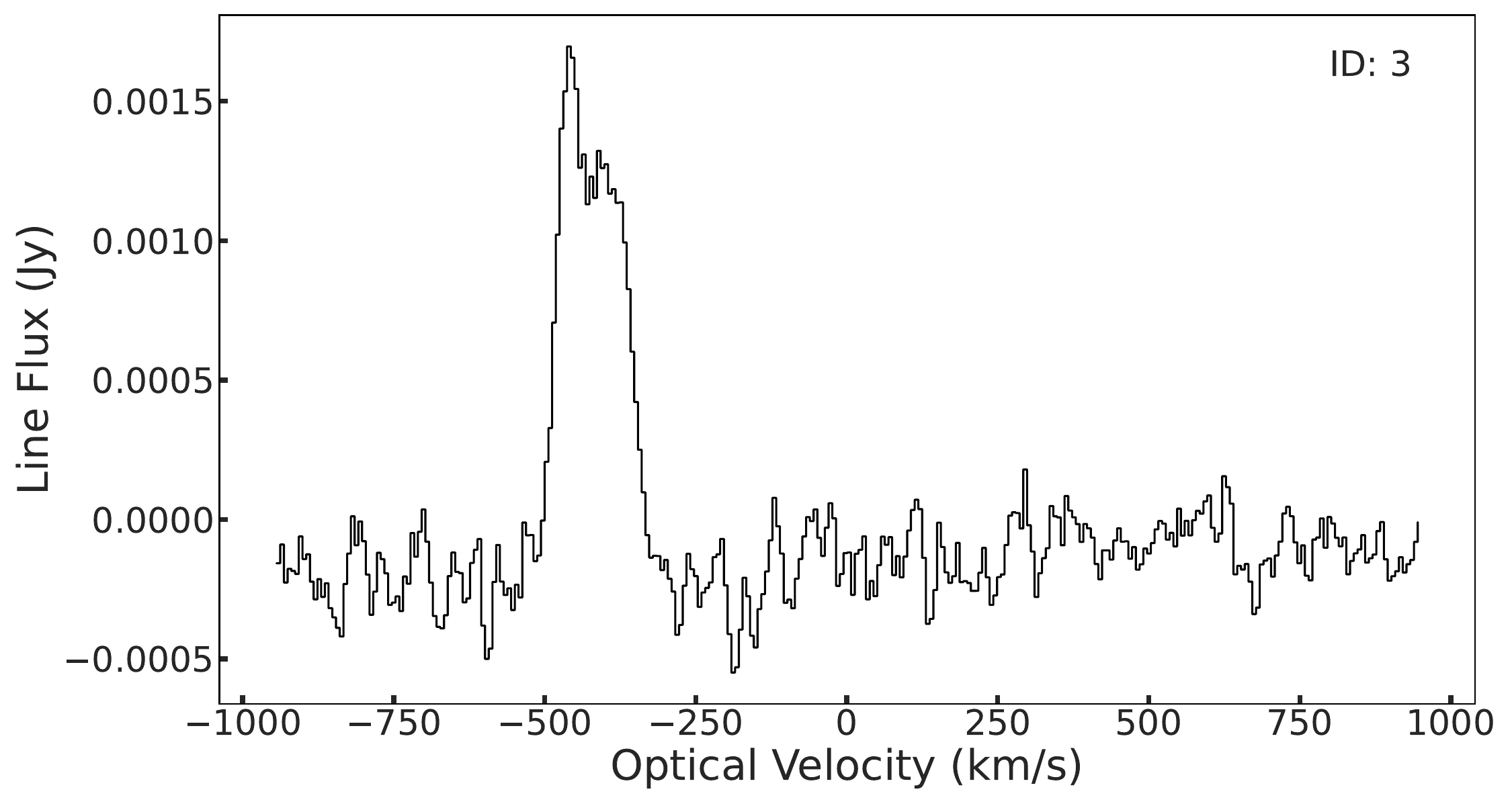}
  \end{subfigure} &
  
  \begin{subfigure}{0.45\textwidth}
    \includegraphics[width=\linewidth]{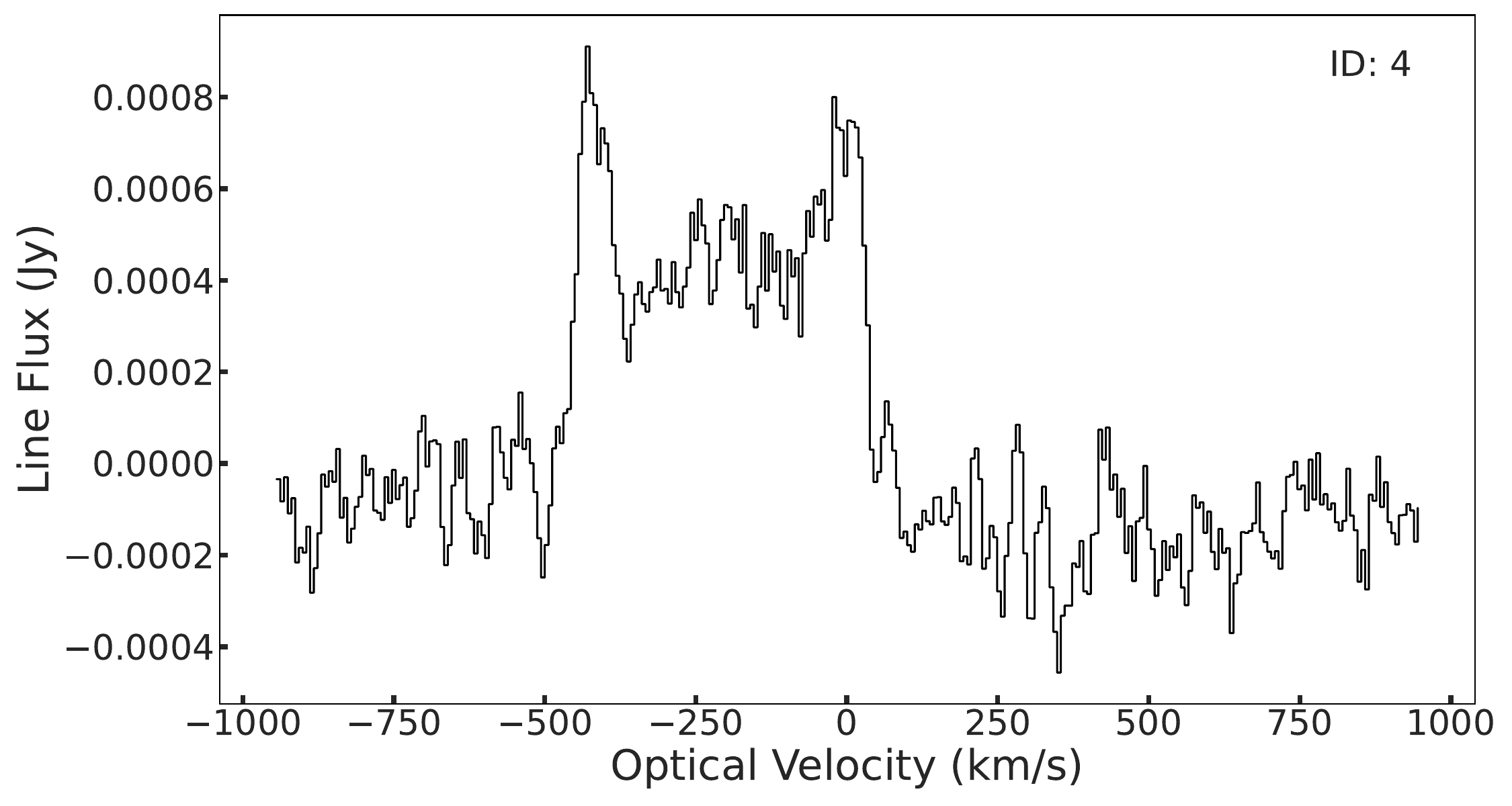}
  \end{subfigure} \\
  
  \begin{subfigure}{0.45\textwidth}
    \includegraphics[width=\linewidth]{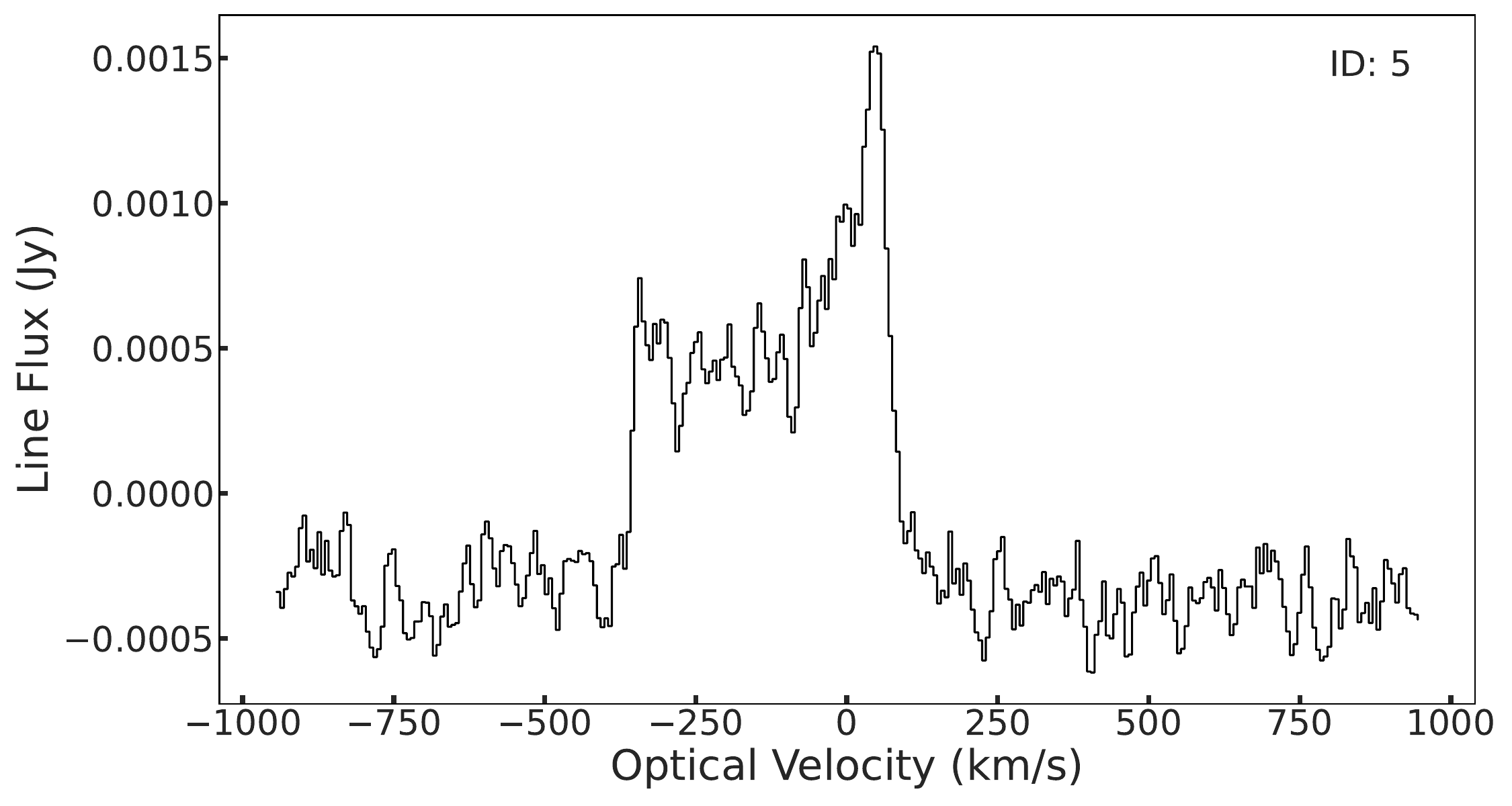}
  \end{subfigure} &
  
  \begin{subfigure}{0.45\textwidth}
    \includegraphics[width=\linewidth]{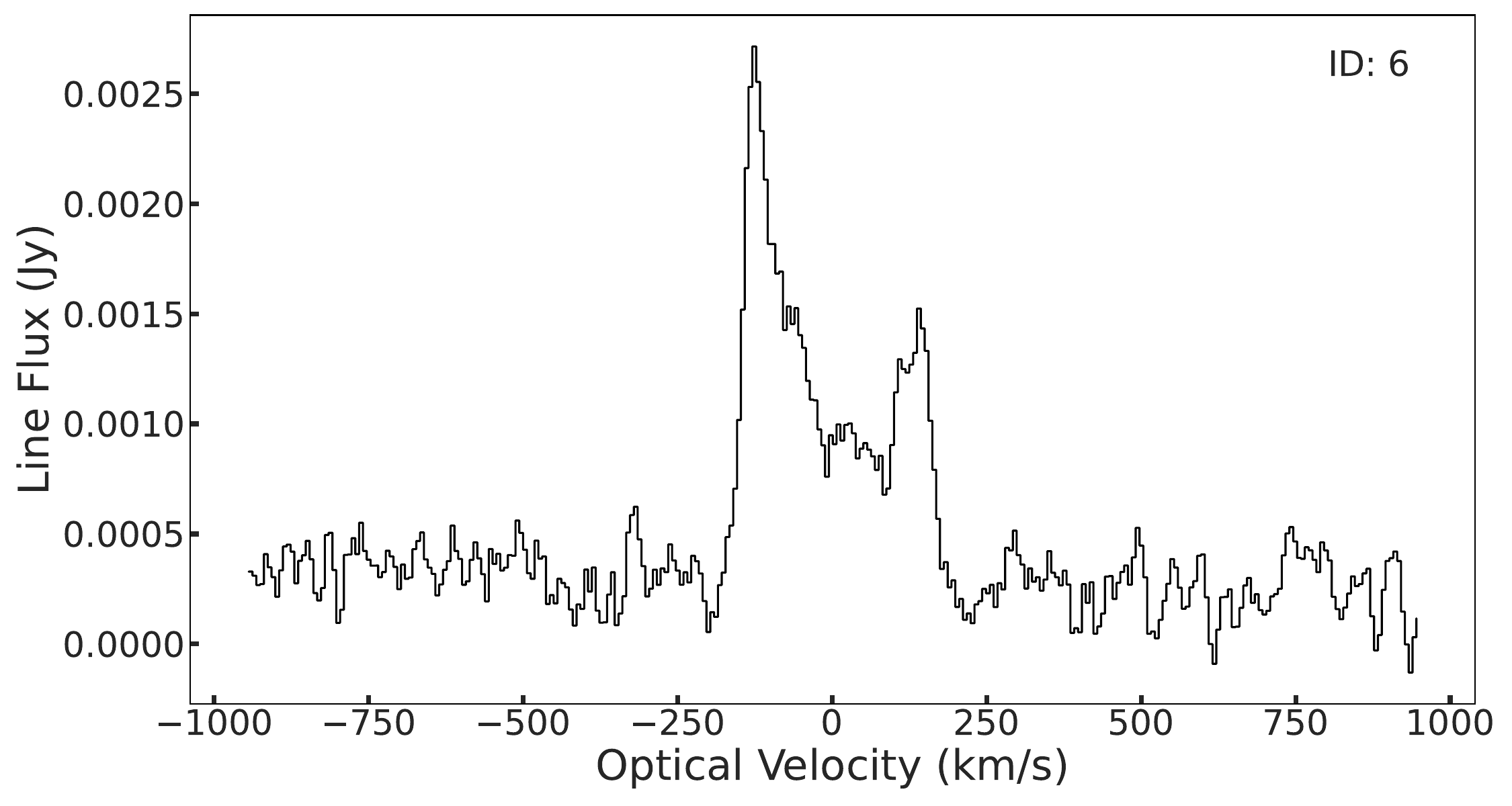} 
  \end{subfigure} \\
  
  \begin{subfigure}{0.45\textwidth}
    \includegraphics[width=\linewidth]{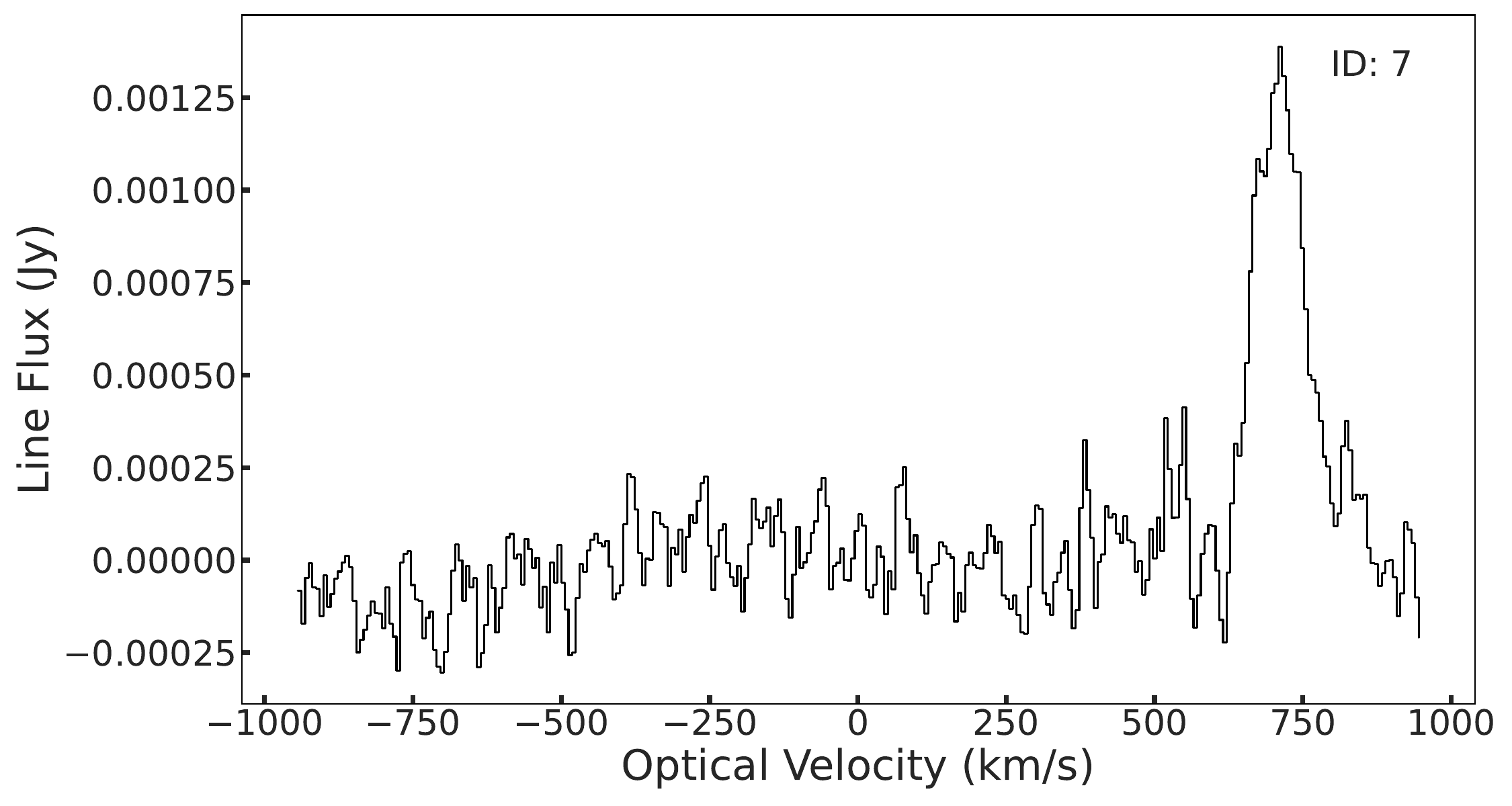}
  \end{subfigure} \\ 
\end{tabular}
\caption{Neutral Hydrogen emission spectra of the seven sources shown in Fig.~\ref{fig:total_intsense_sofia_continuum_contours}. The individual spectra were smoothed using a boxcar filter over three adjacent channels, corresponding to a velocity resolution of 17.6$\, $km$\, $s$^{-1}$.}
\label{fig:emission_spectra}
\end{figure*}

To contextualise the absorption within PKS 2014-55, we also examined the surrounding \hi{} environment. This revealed several sources of \hi{} emission at significantly lower redshifts than PKS 2014-55, potentially indicating the presence of a foreground group. The detection labelled ID 3 in Fig.~\ref{fig:total_intsense_sofia_continuum_contours} is the closest to the host both spectrally and spatially. Although this \hi{} cloud is coincident with the spiral galaxy J2017-5540, its association remains uncertain due to the absence of a recorded optical redshift. The cloud lies 1.1 arcmin (78 kpc) south of PKS 2014-55. Using the \hi{} redshift of the cloud and the 6dF spectroscopic redshift of PKS~2014$-$55, reported as $z = 0.060629 \pm 0.000150$ by \citet{2004Jones}, we estimate a velocity offset of $(0.0053 \pm 0.0002)c \sim 1600 \pm 50$\,km\,s$^{-1}$. This substantial velocity difference effectively rules out the possibility of recent gas exchange between the two systems.\\ \\
The remaining six \hi{} emission detections identified in the field exhibit even larger velocity offsets from that of PKS 2014-55 than ID 3, with velocity separations well in excess of 1000 km~s$^{-1}$. This effectively rules out any of these galaxies as recent gas donors or interaction partners.\\ \\
We estimate the \hi{} column densities and spatial extents for all six emission-line sources detected in the field. As an illustrative example, the \hi{} emission associated with J2017-5540 spans 94 kpc along its longest axis, measured at a column density threshold of $1.3 \times 10^{19}$ cm$^{-2}$. Assuming optically thin emission, we calculate the column densities using Equation 74 from \citet{hi_eqs}:
\begin{equation}
N_{\text{HI}} = 2.33 \times 10^{20} (1 + z)^4 \left( \frac{\text{S}}{\text{JyHz}} \right) \left( \frac{ab}{\text{arcsec}^2} \right)^{-1},
\label{eq:column_density_II}
\end{equation}
where $a$ and $b$ are the beam’s major and minor axes, respectively. The corresponding flux densities ($S_{\nu}$), redshifts ($z$), luminosity distances ($D_{\rm{L}}$), and \hi{} masses ($M_{\rm{HI}}$) for all detections are summarised in Table~\ref{tab:summary_HI_emission}, with their spectra presented in Figure~\ref{fig:emission_spectra}.
\begin{table}
\caption[Main Spectrum Properties]{Absorption spectrum properties.} 
    \centering
    \begin{tabular}{cccc} 
     \toprule[\heavyrulewidth]
         Line                     &  Broad & Deep-Narrow & Shallow\\
    \specialrule{0.1em}{0em}{0em} 
         $S_{\text{peak}}(\text{mJy})$                                  & -1.1 ± 0.1   & -2.6 ± 0.1   & -0.9 ± 0.1 \\
         ${\text{FWHM} \, (\text{km\,s}^{-1})}$                         & 38 ± 7       & 19 ± 6       & 22 ± 6     \\
         $( v_{\rm{sys}} - v_{\rm{peak}}) (\text{km\, s}^{-1})$         & 96 ± 50      & 160 ± 40     & 240 ± 40   \\
         $\tau_{\text{peak}}$                                           & 0.02 ± 0.001 & 0.04 ± 0.002 & 0.01 ± 0.001 \\
         $N_{\text{HI}} (\times 10^{20} \, \text{cm}^{-2})$             & 2.6 ± 0.3    & 3.0 ± 0.2    & 1.3 ± 0.1 \\
    \bottomrule[\heavyrulewidth] 
    \end{tabular}
    \begin{tablenotes}[flushleft]
    \item Notes. $v_{\rm{sys}}$ represents the systemic velocity, while $v_{\rm{peak}}$ indicates the velocity at the peak of the absorption component.
    \end{tablenotes}
    \label{tab:summary_HI}
\end{table}
Taken together, the results suggest that PKS 2014-55 exemplifies a mode of AGN reactivation governed by structured gas inflow, rather than outflow or recent major interactions. The combination of redshifted absorption features, a restarted radio morphology, and the absence of obvious merger indicators provides a compelling case study of cold gas accretion in rejuvenated radio galaxies.
\section{Summary}\label{summary}
To summarise, we have reported new high-spectral resolution (5.8 km$\, $s$^{-1}$) MeerKAT observations of the \hi{} absorption against the central region of the rejuvenated Giant X-shaped radio galaxy PKS 2014-55. The multi-component spectrum shows three features that are redshifted relative to the systemic velocity of the host. Our observations reveal for the first time seven neighbouring \hi{} sources that contain substantial amounts of \hi{} (Fig.~\ref{fig:total_intsense_sofia_continuum_contours}), none of which are close enough to interact with PKS 2014-55. \\ \\ 
The estimated column densities for the three components, from left to right in Fig.~\ref{fig:spectrum}, are $2.6 \pm 0.3$ $\times 10^{20}$, $3.0 \pm 0.2$ $\times 10^{20}$, and $1.3 \pm 0.1$ $\times 10^{20}$\,cm$^{-2}$, assuming $T_{\rm{s}} = 100$\,K, $f_{\rm{c}} = 1$, and $\Delta \nu = 100$\,km\,s$^{-1}$. In a 50-MHz wide bandwidth, we found no blue-shifted wings or other absorption structures.\\ \\
We surmise that the gas seen in absorption is likely fueling the AGN, contributing to its recent reactivation. The detection of redshifted \hi{} in this rejuvenated radio galaxy supports the growing view that cold gas accretion, as traced through associated absorption, plays an important role in triggering renewed AGN activity. PKS 2014-55 adds to the small but increasing number of powerful radio galaxies in which \hi{} absorption offers direct observational evidence of nuclear fueling during phases of episodic activity.\\ \\
Future high-resolution observations are essential to localise the \hi{} absorption and better constrain the physical conditions of the infalling gas. Observations of CO lines would offer complementary insight into the molecular component, potentially revealing a multiphase cold gas structure. Both objectives can be addressed with targeted ALMA follow-up. When considered alongside its restarted morphology, PKS 2014-55 emerges as a promising testbed for exploring how episodic accretion events may trigger recurrent radio activity. 
\begin{table*}
    \centering
    \scriptsize
    \caption[Main Spectrum Properties]{Positions, integrated flux densities, W50s, redshifts, luminosity distances, and masses of the \hi{} emission in the PKS 2014-55 group.}
    \begin{tabular}{@{} c c c c c c c c c @{}} 
        \toprule[\heavyrulewidth]
        \textbf{ID} & \textbf{\(\alpha\)} & \textbf{\(\delta\)} & \textbf{\(S_{\nu} \pm \delta S\) [\(\times 10^2\)]\,JyHz} & \textbf{\(W_{50}\) [\(\times 10^5\)]\,Hz} & \textbf{z}$^{(\dag)}$ & \(D_\text{L}^{(\ddag)}\) [Mpc] & \textbf{\(M_{\text{HI}} \pm \delta M\) [\(\times 10^9\)] \(M_\odot\)} & \textbf{\(N_{\text{HI}} \pm \delta N\) [\(\times 10^{20}\)]\,cm$^{-2}$} \\
        \specialrule{0.1em}{0em}{0em}
        1 & 20:18:11.68 & -55:44:49.7 & 3.4 ± 0.4  & 6.66 & 0.0514 & 229 & 0.88 ± 0.10 & 1.07 ± 0.10 \\
        2 & 20:18:10.05 & -55:37:11.4 & 3.2 ± 0.3  & 1.75 & 0.0550 & 245 & 0.96 ± 0.08 & 1.03 ± 0.07 \\
        3 & 20:17:57.17 & -55:40:37.5 & 13.3 ± 0.6 & 5.33 & 0.0553 & 247 & 4.0 ± 0.2  & 4.26 ± 0.16 \\
        4 & 20:17:46.75 & -55:36:16.9 & 14.9 ± 0.8 & 20.4 & 0.0561 & 250 & 4.7 ± 0.2  & 4.81 ± 0.20 \\
        5 & 20:16:49.25 & -55:38:49.7 & 20.3 ± 1.1 & 18.0 & 0.0562 & 251 & 6.4 ± 0.3  & 6.5 ± 0.3 \\
        6 & 20:17:58.96 & -55:33:12.4 & 40.3 ± 1.4 & 13.4 & 0.0565 & 253 & 12.8 ± 0.4 & 13.0 ± 0.4 \\
        7 & 20:18:45.97 & -55:47:24.3 & 7.9 ± 0.5  & 3.95 & 0.0587 & 262 & 2.7 ± 0.2  & 2.57 ± 0.12 \\
        8$^{(*)}$ & 20:18:01.27 & -55:39:29.6 & 21.2 ± 0.8 & 2.14 & 0.0571 & 255 & 6.9 ± 0.3  & 6.84 ± 0.20 \\
        \bottomrule[\heavyrulewidth]
    \end{tabular}
    \vspace{0.5em}
    \begin{tablenotes}[flushleft]
    \item Notes. $^{(\dag)}$ We estimate the uncertainty in the \hi{}-redshift to be $\pm$1.84e-05. $^{(*)}$ Denotes absorption detection. $^{(**)}$ Column density as a function of brightness temperature. $^{(\ddag)}$The luminosity distance uncertainty is 0.0788 Mpc.
    \end{tablenotes}
    \label{tab:summary_HI_emission}
\end{table*}
\section*{Acknowledgements}
We acknowledge the use of the ilifu cloud computing facility \url{www.ilifu.ac.za}, a partnership between the University of Cape Town, the University of the Western Cape, Stellenbosch University, Sol Plaatje University, the Cape Peninsula University of Technology, and the South African Radio Astronomy Observatory. The ilifu facility is supported by contributions from the Inter-University Institute for Data Intensive Astronomy (IDIA –a partnership between the University of Cape Town, the University of Pretoria, and the University of the Western Cape), the Computational Biology division at UCT, and the Data Intensive Research Initiative of South Africa (DIRISA). \\ \\
This work was carried out using the data processing pipelines developed at the Inter-University Institute for Data Intensive Astronomy (IDIA) and available at https://idia-pipelines.github.io. IDIA is a partnership of the University of Cape Town, the University of Pretoria, and the University of the Western Cape.\\ \\
This work made use of the CARTA (Cube Analysis and Rendering Tool for Astronomy) software (DOI 10.5281/zenodo.3377984 –  https://cartavis.github.io).\\ \\
(Part of) the data published here have been reduced using the $\bf{{\tt CARACal}}$ pipeline, partially supported by ERC Starting grant number 679627 'FORNAX', MAECI Grant Number ZA18GR02, DST-NRF Grant Number 113121 as part of the ISARP Joint Research Scheme, and BMBF project 05A17PC2 for D-MeerKAT. Information about $\bf{{\tt CARACal}}$ can be obtained online under the URL: \url{https://caracal.readthedocs.io}.\\ \\
The MeerKAT telescope is operated by the South African Radio Astronomy Observatory, which is a facility of the National Research Foundation, an agency of the Department of Science and Innovation.\\ \\
The National Radio Astronomy Observatory is a facility of the National Science Foundation, operated under a cooperative agreement by Associated Universities, Inc.\\ \\
FMM acknowledges that part of the research activities described in this paper were carried out with contribution of the Next Generation EU funds within the National Recovery and Resilience Plan (PNRR), Mission 4 - Education and Research, Component 2 - From Research to Business (M4C2), Investment Line 3.1 - Strengthening and creation of Research Infrastructures, Project IR0000034 – “STILES - Strengthening the Italian Leadership in ELT and SKA.
\section*{Data Availability}
The raw data is available from the SARAO archive (\url{https://archive.sarao.ac.za}) under project code SCI-20180624-FC-01. Final image cubes can be obtained from \url{https://zenodo.org/records/16929331}.
\bibliographystyle{mnras}
\bibliography{References}
\bsp
\label{lastpage}
\end{document}